\documentclass[11pt]{article}
\usepackage{fleqn,cospar}

\usepackage{epsf}
\usepackage[figuresright]{rotating}

\newcommand{\Rin}{R_{\rm in}}
\newcommand{\Rg} {R_{\rm g}}
\newcommand{\Rms} {R_{\rm ms}}
\newcommand{\NH} {N_{\rm H}}
\newcommand{\MSun} {{\rm M}_{\odot}}
\newcommand{\kT} {k T_{\rm e}}
\newcommand{\Ptot} {P_{\rm tot}}
\newcommand{\tdyn} {t_{\rm dyn}}
\newcommand{\tel} {\tau_{\rm es}}
\newcommand{\hi} {\chi^2_{\nu}}
\newcommand{\Ka}{K$_{\alpha}$}

\title{X--RAY ENERGY AND POWER SPECTRA OF SEYFERT 1 AND NARROW LINE 
SEYFERT  1 GALAXIES: COMPARISON TO BLACK HOLE BINARIES}

\author{P. T. \.{Z}ycki\address{Nicolaus Copernicus Astronomical Center,
    Bartycka 18, 00-716 Warsaw, Poland, (ptz@camk.edu.pl)}}

\begin{document}

\maketitle

\begin{abstract}
I review observational data on X--ray energy spectra and variability
of Seyfert 1 and Narrow Line Seyfert 1 galaxies, and compare them
to black hole binary (BHB) systems. There seems to be greater similarity
of time averaged properties (i.e.\ energy spectra) than time variability
between the supermassive and stellar-mass black hole accreting systems.

Energy spectra of Seyfert 1 galaxies are similar to BHB in low/hard state,
but power spectra of Sy 1 are intermediate in properties between those
for hard and soft states of BHB.

There is growing evidence that NLSy 1 are accreting at higher accretion
rate than Sy 1, which supports the view that they correspond to soft states
of BHB. Again, they are more similar to their stellar counterparts with
respect to their spectral properties than to timing properties.

I briefly outline some theoretical ideas which suggest explanations
of those differences.

\end{abstract}

\section*{SEYFERT 1 GALAXIES}

\subsection*{Energy spectra}

\subsubsection*{Primary continua}

Broad band X--ray/$\gamma$--ray spectra of Seyfert 1 galaxies are
compatible with Compton upscattering of soft seed photons by energetic,
thermal electrons (see reviews in Mushotzky et al.\ 1993;
Zdziarski 1999).

Spectral slopes span a range of values, $\Gamma = 1.6$--2.0, the exact
limits being primarily instrument-dependent, but the intrinsic dispersion
of parameters does seem to be real (Nandra et al.\ 1997). Individual values
of electron optical depth, $\tel$, and temperature, $\kT$, are not
reliably estimated due to a number of factors. First attempts used
incorrect analytical approximations to Comptonized spectra, which 
overestimated the temperature and, as a consequence, gave low optical
depth, $\tel \sim 0.1$. Later, better models gave $\kT \approx 100$ keV
and $\tel\sim 1$ (exact values are obviously geometry-dependent).
Recent simultaneous {\it ASCA-RXTE-OSSE\/} data of IC~4329a yield 
$\kT \approx 100$ keV (Done et al.\ 2000). On the other hand, a number
of {\it BeppoSAX\/} observations suggest rather higher temperatures (see
Matt 2000 for review). In particular, data from NGC~5548 prefer
$\kT > 200$ keV . These data are compatible with various
geometries of the Comptonizing cloud (slab, hemisphere, sphere). Fitting
detailed spectral model for each geometry yields different values of 
$\tel$, $\kT$ and reflection amplitude (Petrucci et al.\ 2000).

\subsubsection*{The X--ray reprocessed component}

\begin{figure}
\epsfysize = 5 cm
\hfil\epsfbox[18 400 600 700]{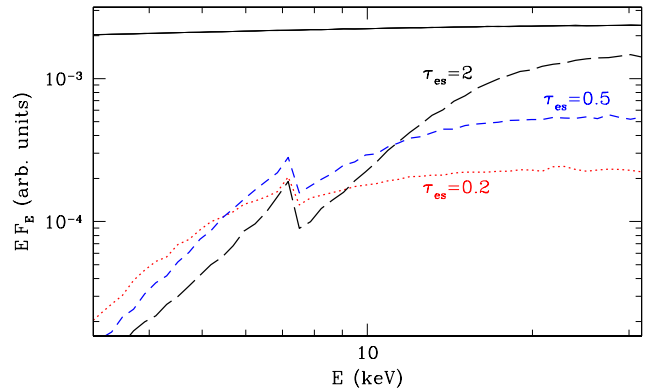}\hfil
\caption{Compton-reflected continua from tori with opening (funnel) half-angle
$40^{\circ}$, observed at $i=30^{\circ}$, for different optical thickness
of the torus (square cross-section of the torus is assumed).
Although the ``reflection hump'' near 20--40 keV is suppressed when
$\tel < 1$, the continuum below 10 keV is not. The Fe \Ka\ line EW is $>50$ 
eV for $\tel \ge 0.2$.
\label{zycki_fig:thin}
}
\end{figure}

The spectral component (Compton-reflected continuum and Fe fluorescent
\Ka\ line)
due to reprocessing by cool, optically thick
plasma is ubiquitously present in Sy 1 spectra. There is increasing
evidence that it consists of two components: one coming from an accretion
disk close to the central BH and X--ray source, the other coming from
a distant reprocessor, most likely the molecular torus invoked
in AGN  unification models. The evidences come from many 
directions: the most direct one are the {\it BeppoSAX\/} observations of 
NGC 4051, when the central X--ray source "switched off" and the only
detectable emission came in the form of pure reprocessed component
(Guainazzi et al.\ 1998).
Additional  narrow Fe \Ka\ line component was detected in some sources
through direct modeling of good quality data 
(MCG~5-23-16, Weaver et al.\ 1997; 
IC~4329a; Done et al.\ 2000; see also Lubi\'{n}ski \& Zdziarski 2000). 
Interestingly, Lubi\'{n}ski \& Zdziarski (2000) claim that the narrow
\Ka\ line in average ASCA spectra is not accompanied by the 
Compton-reflected continuum,
i.e.\ the reprocessing material is not Thomson thick. However,
even if the reprocessor is optically thin(ish), the K shell absorption
edge should be detectable in the data. The measured equivalent width 
of the narrow line, $\approx 50$ eV, requires 
$\NH\approx 2.5\times10^{23}\,{\rm cm^{-2}}$,
i.e.\ $\tau_{es}=0.2$,
for the torus half-opening (funnel) angle of $40^{\circ}$.
Such a reprocessor would produce the reflected continuum which,
below 10 keV, is {\em not\/} suppressed compared to the usual,
optically thick reprocessing  (Fig.~\ref{zycki_fig:thin}).

Yet another evidence for two reprocessors comes from considerations 
involving X--ray variability: the reprocessed component seems to 
be generally less variable than the primary continuum (Reynolds 2000; 
Chiang et al.\ 2000; Done et al.\ 2000). A contribution from a distant,
hence $\sim$constant component makes this fact easier to understand.

Of paramount importance for constraining theoretical models is
the amplitude (and other parameters) of the disk--reprocessed component.
Its amplitude is however uncertain, since the spectral decomposition
into the two components is not unique. In NGC 4051 the solid angle
of the distant torus seems to be quite large, $\Omega/2\pi \sim 0.5$, 
which should be then
subtracted from the total amplitude, if this value is universal. 
If the torus is Thomson thin, no correction  is necessary, provided
that the amplitude is determined from fitting broad band spectra. 
However, even without the correction the
data suggest that the amplitude, at least in some sources, 
is $<1$, where $R=1$ ($\Omega=2\pi$) corresponds to an
isotropic source above a cold (i.e.\ metals not ionized), flat, infinite 
disk (Magdziarz et al.\ 1998; Chiang et al.\ 2000; Done et al.\ 2000; 
Lubi\'{n}ski \& Zdziarski 2000). 
One source where the amplitude is certainly $\approx 1$ is
MCG --6-30-15 (Lee et al.\ 1999).

The Fe spectral features superimposed on the reflected continuum are
not sharp, but broadened and smeared (Tanaka et al.\ 1995).
The standard interpretation is that the reprocessing takes place
in the accretion disk, close to the central black hole, so the smearing
is due to Doppler effect and gravitational redshift.
Broadening by Comptonization in transmission can indeed be
ruled out (see Reynolds and Wilms 2000 and references therein), 
but contribution from Comptonization in reflection is a possibility 
(Karas et al.\ 2000). The inner radius of the reprocessing disk,
as inferred from the smearing, is usually larger than the last
stable orbit ($\Rms = 6 \Rg = 6 GM/c^2$ in Schwarzchild metric), for
the irradiation emissivity $\propto r^{-3}$.
The best known case of very broad line in MCG--6-30-15 (Tanaka et al.\ 1995;
Iwasawa et al.\ 1996) seems to be an exception rather than a rule. Analysis of
{\it ASCA\/} data (Nandra et al.\ 1997) already showed a range of the
smearing parameters. Detailed analysis of {\it ASCA}--{\it RXTE\/} data
of IC~4329a (Done et al.\ 2000) gives the inner radius $\Rin = 15$--$100\Rg$,
depending on Fe abundance and source inclination. Similar values are obtained
for NGC~5548 (Chiang et al.\ 2000 and J.~Chiang, private communication).
Re-analysis of {\it ASCA\/} data by Lubi\'{n}ski \& Zdziarski (2000)
gives $\Rin$ compatible with $6\,\Rg$ for their softer sub-samples of
objects, but $\Rin > 6\,\Rg$ at high significance for the hard sub-sample.

Of certain importance for correct, quantitative determination of $\Rin$
from relativistic smearing is using models that apply the smearing to
both the Fe \Ka\ line and absorption edge (\.{Z}ycki et al.\ 1997). 
The decomposition of counts
into spectral features and continuum crucially depends on details of the
shape of the model.

\subsubsection*{Correlations between parameters}

A number of correlations between spectral parameters is observed.

\smallskip
{\it Sample of objects.}\ In a sample of {\it Ginga\/} data (Seyfert 1 
galaxies, 
X--ray binaries with both BH and NS central objects) analysed homogeneously
by Zdziarski et al.\ (1999) a correlation was found between the spectral slope
of the primary power law,
$\Gamma$, and the amplitude of the reprocessed component, $R$. Softer 
spectra (larger $\Gamma$) contain stronger reprocessed components.
The correlation appears to be very significant, even correcting for
the intrinsic  $R$--$\Gamma$ correlation in the data fitting procedure.
Individual values of $R$ and $\Gamma$ are not always reliably estimated, 
especially
at the extremes of the correlation, e.g.\ $R$ for MCG--6-30-15 from broad 
band {\it RXTE\/} data is $\approx 1$ (Lee et al.\ 1999) rather than 
$\approx 2$  as in Zdziarski et al.\ (1999). 
The correlation is also seen in the {\it BeppoSAX\/}
sample of Seyfert 1 galaxies (Matt 2000) as well as in
a sample of {\it ASCA\/} data (Lubi\'{n}ski \& Zdziarski 2000). 

Moreover, a correlation between the spectral width
of the Fe \Ka\ line and $\Gamma$ (or $R$) is observed: the line is broader
when the spectrum is steeper and $R$ larger (Lubi\'{n}ski \& Zdziarski 2000;
Zdziarski, these proceedings).

\smallskip
{\it Time histories of individual objects.}\ The $R$--$\Gamma$ correlation
was also observed in time histories of individual objects. Significant 
correlation ($>99\%$) was found in series of {\it Ginga\/} observations 
of NGC~5548 (Magdziarz et al.\ 1998), although
Chiang et al.\ (2000) report much weaker correlation in {\it RXTE\/} data, 
formally consistent with constant $R$ for $\Gamma$ changing by 0.2.
The correlation  was observed in {\it Ginga\/} data of Nova Muscae 1991
(\.{Z}ycki et al.\ 1998ab), during the source's decline
after outburst. It is also seen in GX~339-4, both in
{\it Ginga\/} data (Ueda et al.\ 1994; Zdziarski et al.\ 1999) and 
{\it RXTE\/} data (Revnivtsev et al.\ 1999b, Gilfanov et al.\ 2000b).

The fact that the $R$--$\Gamma$ correlation seems to be present both
in a sample of objects and in time histories of individual objects
is not a trivial one. Variations of parameters between different
objects in a sample are not necessarily caused by the same mechanism
as time variations of parameters in a history of a given object.

\subsection*{Time variability}

\subsubsection*{Power density spectra}

Power density spectra (PDS) of X--ray variability show generally
power law shapes, $P(f) \propto f^{-\alpha}$ , with slopes $\alpha = 1$--2
(Lawrence et al.\ 1987; McHardy \& Czerny 1987).
Breaks (flattening to $f^{-\alpha}$ with $\alpha\approx 0$) in the PDS 
are thus expected at low $f$ to avoid divergent power.
First attempts to find the breaks  used {\it EXOSAT\/} long--look
data, where a break was reported in NGC~5506 near $10^{-7}$ Hz (McHardy 1989).
Further cases were found in {\it Ginga\/} data (Hayashida et al.\ 1998
and references therein).
{\it RXTE\/} campaigns brought good quality data which enabled determination
of PDS for a number of sources. Clear break at $f\sim 10^{-7}$--$10^{-6}$ Hz 
is seen in PDS of NGC~3516, although it precise position and shape is 
uncertain (Edelson \& Nandra 1999). A break at $f \sim 10^{-7}$ Hz is suggested
by the {\it RXTE}-ASM data of NGC~5548 (change of slope from $f^{0}$ to 
$f^{-1}$),
with further steepening to $f^{-2}$ at $f\sim 10^{-6}$ Hz, when {\it RXTE}-PCA 
and {\it ASCA\/} data added (Chiang et al.\ 2000).
From {\it RXTE}-PCA and {\it ASCA\/} data alone a break at 
$f \sim 10^{-5}$ Hz is seen in  MCG--6-30-15 (Nowak \& Chiang 2000).

Integrated r.m.s.\ amplitude of variability is 20\%--30\% (Edelson 2000).

\subsubsection*{Time lags}

Time lags between different energy bands are expected if the X--rays
are indeed produced by inverse-Compton process. 
However, no significant time lags between the hard, {\it RXTE}-PCA band and
the soft, {\it ASCA\/} band have been detected so far.
In MCG --6-30-15 the lag between the 8--15 keV {\it RXTE\/} band and the soft,
0.5--1 keV {\it ASCA\/}  band is consistent with 0, with the 90\% upper limit 
of 2 ksec (Nowak \& Chiang 2000). 
Similarly, no time lag was detected between the 0.5--2 keV {\it ASCA\/} band 
and {\it RXTE}-PCA, 2--10 keV band in NGC~3516 (Edelson et al.\ 2000).
In NGC~5548 the lag between the 2--20 keV {\it RXTE}-PCA band and the 
0.5--1 keV {\it ASCA\/} band is $5^{+7}_{-9}$ ksec (99\% confidence limits). 
Significant lags were detected in NGC~5548 between the 
{\it RXTE}-PCA band and the {\it EUVE\/} band (around 0.1 keV) of 
$34^{+25}_{-18}$ ksec, and between the {\it ASCA\/} band and the
{\it EUVE\/} band by $13^{+11}_{-11}$ ksec, both being 99\% 
confidence limits (Chiang et al.\ 2000).

\subsection*{Comparison to black hole binaries}

\subsubsection*{Energy spectra}

There is a general similarity between hard X--ray spectra of Seyfert 1s
and BHB in low/hard state. Both classes show a range of spectral indices,
amplitudes of reflection and relativistic smearing of Fe spectral features.

Individual, well-studied cases of BHB include
Cyg X-1 (Poutanen et al.\ 1997; Gierli\'{n}ski et al.\ 1997; Dove et al.\
1997; Done \& \.{Z}ycki 1999; Di Salvo et al.\ 2000), 
GX~339-4 (Ueda et al.\ 1994; Zdziarski et al.\ 1998; Wilms et al.\ 1999),
LMC X-3 (Boyd et al. 2000). 
Results are reviewed and summarized in Poutanen (1998)
and Done (these proceedings).

The observed, hard ($\Gamma < 2$) spectral slopes indicate photon-starved 
plasma, compared to uniform, static disk-corona models 
(Haardt \& Maraschi 1991). At least three models 
can explain the observed range of parameters and correlations
(see Di Salvo et al.\ 2000 for recent discussion and application 
to Cyg X--1): truncated 
standard disk with inner hot flow (Esin et al.\ 1997;
Meyer, Liu \& Meyer-Hofmeister 2000; R\'{o}\.{z}a\'{n}ska \& Czerny 2000), 
relativistic plasma ejection
from magnetic flares (Beloborodov 1999) and accretion disk with a "hot skin" 
(Nayakshin et al.\ 2000; \.{Z}ycki \& R\'{o}\.{z}a\'{n}ska 2000).
However, only the first one naturally explains the general absence
of the broadest (most strongly relativistically smeared) component of 
the reprocessed
spectrum, i.e.\ the fact that in most spectra the inner disk radius,
$\Rin$, as inferred from fitting the shape of the Fe \Ka\ line and edge
is larger than the marginally stable orbit. A rather dramatic change of
structure of the disk--active corona system at $r \le \Rin$ 
would be required to remove the broadest components in the other models
(e.g.\ thickness of the ``hot skin'' $\tau_{\rm hot} \gg 1$ for $r<\Rin$), 
while the predicted radial dependencies do not show 
any discontinuities (\.{Z}ycki \& R\'{o}\.{z}a\'{n}ska 2000).
Since the inferred values
of $\Rin$ are similar in Sy 1 and BHB (20--100 $\Rg$), the relevant mechanism
should be at most only weakly dependent on central mass. This excludes e.g.\ 
radiation pressure instabilities which affect much more strongly the
$\alpha\Ptot$ SS disks around supermassive 
black holes than the stellar mass ones.
On the other hand, $\Rin$ predicted  by the disk evaporation model of
R\'{o}\.{z}a\'{n}ska \& Czerny (2000) are independent of the central mass.
Presence of a characteristic radius is also supported by the correlated
spectral--temporal behavior observed in e.g.\ Cyg X--1, where correlations
are seen between the spectral slope, amplitude of the reprocessed 
component and the peak of the PDS (Gilfanov et al.\ 1999).

\subsubsection*{Time variability}

\begin{figure}
\epsfysize = 7 cm
\hfil\epsfbox[20 240 600 700]{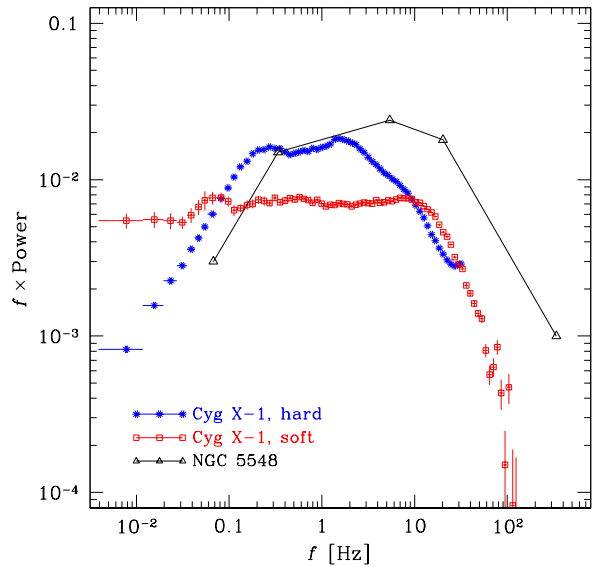}\hfil
\caption{Comparison of X--ray PDS of Cyg X--1 in two spectral states with 
the X--ray PDS of NGC~5548 ({\it RXTE\/} data). The PDS of NGC~5548 was shifted
in $\log f$ by 6.8, derived from the best available estimate of the mass ratio
of central BHs. The position of the high-$f$ break in PDS (it is this break 
that
seems to scale with the BH mass) in NGC~5548 corresponds to the {\em soft\/}
state of Cyg X--1, rather than the hard state. From Czerny et al.\ (2000)
(Cyg X--1 data from Gilfanov et al.\ 2000a).
\label{fig:psdcygx1}}
\end{figure}

{\it PDS.}\  
Observations of time variability of BHBs are much easier than AGNs on 
almost all time scales except for the dynamical time scale, 
$\tdyn = \Omega_{\rm K}^{-1} \approx 2\times 10^{-4}\, M/\MSun$ sec. 
Therefore positions of the breaks in the PDSs 
of the former objects were determined more accurately, and were used
to estimate BH masses in AGNs, whenever a break in PDS of an AGN was
found. In a variant of the method, Hayashida et al.\ (1998) used 
the frequency at which the value of PDS ($({\rm r.m.s.})^2\,{\rm Hz}^{-1}$)
is $10^{-3}$. An implicit assumption here is
obviously that the variability mechanisms are the same in both classes
and the only dependence is through the central mass (i.e.\ the characteristic
size measured in gravitational radii is the same). With more reliable 
independent mass estimates of central BH in Sy 1s, it became possible to
use an AGN PDS as the reference, and to {\it test\/} the implicit assumption 
of the previous method. This has been done by Czerny et al.\ (2000)
who adopted NGC~5548 as the reference object. Central BH mass estimated 
by a number of methods is close to $7\times 10^7\MSun$, while its PDS 
was determined thanks to {\it RXTE\/} monitoring campaign 
(Chiang et al.\ 2000). 
Comparison to Cyg X--1 PDS in both the low/hard and soft
states (data from Gilfanov et al.\ 2000a)
reveals important differences between them, both in the position of 
the break and the overall normalization. Frequency-integrated r.m.s.\ 
variability is larger for NGC~5548 than for either spectral state of Cyg X-1
(NGC~5548: 35\%, Cyg X-1, hard state: 28\%, Cyg X-1, soft state 24\%).
Even more importantly, Czerny et al.\ (2000) draw attention to the fact 
that it is the higher-$f$ break 
(where the slope changes to $-2$), which is more relevant for mass
comparison, since the lower-$f$ break is luminosity dependent 
(e.g.\ Belloni \& Hasinger 1990).
 Position of the high-$f$ break in NGC~5548 agrees with similar break
in the {\em soft\/} state of Cyg X--1 rather than in the hard state PDS (
see Fig.~\ref{fig:psdcygx1}).

{\it Time lags.}\ The issue of time lags between different X-ray bands 
has received a lot of attention
during the last couple of years, mainly thanks to a wealth of new data
from {\it RXTE.} The lags were determined for many BHB over a broad 
range of Fourier frequencies 
(e.g.\ Cyg X--1: Miyamoto et al.\ 1988; Cui et al.\ 1997; Nowak et al.\ 1999a;
GX~339-4: Nowak et al.\ 1999b).
They range from almost 0.1 sec at $f\approx 0.1$ Hz to $\sim 10^{-3}$ sec 
at $f \ge 10$ Hz (exact values obviously depend on energy bands).
The lags seem to scale with the mass of central BH in roughly the
same way as the PDS do (see Nowak \& Chiang 2000 for MCG --6-30-15).
The time lags are surprisingly long: at the frequency where the PDS peaks, 
the lags
are  $\sim 100$ times longer than the light crossing time scale of
the region, where most of the gravitational energy is dissipated.

{\it Spectral--temporal correlations.}\ Important correlations are found
between spectral and temporal parameters. In the hard state of Cyg X--1 
the reprocessed component does {\em not\/} respond to primary
continuum variability on time-scales shorter than $\sim 0.5$ sec 
(Revnivtsev et al.\ 1999a; Gilfanov et al.\ 2000a). The same seems to
be true for Sy 1s: as already mentioned, a number of {\it RXTE\/} observations
of Sy 1 show the reprocessed component to be surprisingly weakly variable
on time scales of $\sim$day.

\subsubsection*{X--ray spectral slope -- flux relations}

\begin{figure}
\epsfysize = 5 cm
\hfil\epsfbox[18 400 600 700]{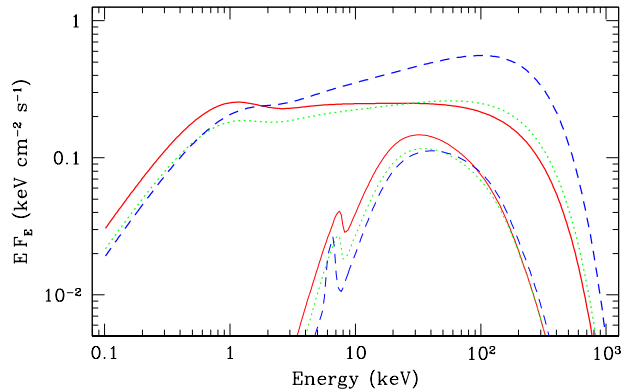}\hfil
\caption{Broad band spectra of Nova Muscae 1991 in low/hard state
({\it Ginga\/} 1--20 keV data only, Comptonization $\kT=100$ keV assumed). 
Clear hardening
of the spectrum with increasing X--ray flux is seen, opposite to what is
observed in e.g.\ Sy 1 galaxy IC~4329a (Done et al.\ 2000).
\label{zycki_fig:nm}
}
\end{figure}

A variety of behaviour is observed here. {\it RXTE\/} data of IC~4329a 
(Done et al.\ 2000; Madejski et al.\ these proceedings) show softening of 
the spectrum when the source
brightens on $\sim$days time scale. The same correlation is seen
on very long time scale (months for BHB, i.e.\ not observable for AGN),
when SXT decline from outburst, from soft spectra at high luminosities
to hard spectra at low luminosities. It is thus interesting to note
that during the gradual decline of Nova Muscae 1991, when the source
was already in low/hard state, the gradual 
hardening of the X--ray slope (\.{Z}ycki et al.\ 1998ab), was accompanied
by an {\em increase\/} of the source X--ray luminosity 
(so called re-flare; Fig.\ref{zycki_fig:nm}). 
Yet another example is the low/hard state of Cyg X--1, where the source 
showed almost stable spectral shape even though its hard X--ray luminosity 
seemed to
have changed by a factor of $\sim 4$ (Gierli\'{n}ski et al.\ 1997).

\subsubsection*{Discussion}

X--ray sources in Seyfert 1 galaxies seem to be more similar to their
small mass counterparts in BHB with respect to their energy spectral
characteristic, than to the temporal behaviour. As the energy spectra are
``time envelopes'' of the inverse-Compton upscattering process (Hua \&
Titarchuk 1996),
they may, as such, correspond to a number of different temporal characteristics
(see \.{Z}ycki 2000 for another example).
Both aspects (and correlations between them)  need to be studied carefully 
in order to discriminate between models.

A number of specific models for generating X--ray light curves with observed 
properties were proposed (see Poutanen 2000 for review). 
Two of the models seem to be most promising: the magnetic flares model 
(Poutanen \& Fabian 1999) and the drifting blobs model 
(B\"{o}tcher \& Liang 1999).
The former model fits the scenario invoking magnetically active corona
as the source of hard X--rays, whilst the latter fits the scenario of
truncated disk with inner hot flow. Both models can in principle be applied to
both BHB and Seyfert 1a, although the discussion so far concentrated
on the BHB case. With both scenarios the light curves are modeled as
superpositions of individual flares, with possible correlations between
the flares (``avalanches'') to account for the long time scale where the PDS
peaks. This approach is vindicated by e.g.\ results
of non-linear prediction method analysis of light curves, demonstrating
that a stochastic type of process is favoured as the origin of X--ray
variability (Czerny \& Lehto 1997). Both models explain the hard X--ray
lags as a result of spectral evolution (hardening) during the flares, 
and so remove
the need for a very spatially extended X--ray source ($\sim 10^4\Rg$),
necessary, if the lags were simply interpreted as photon diffusion time.

Possible differences in time variability between Sy 1 and BHB may be related
to different contribution of radiation pressure to total pressure in accretion
disks around supermassive and stellar mass BH, in connection with the
disk evaporation model of R\'{o}\.{z}a\'{n}ska \& Czerny (2000). They show
that the unstable branch in the $\dot m$--$\Sigma$ diagram is much less
pronounced in BHB than in AGN, if the evaporation branch is added to the
solution. Perhaps the disks in BHB are stable and the entire variability
comes from properties of the Comptonization process 
(Churazov et al.\ 2000),
but the unstable disks in Sy 1 cause additional variability of the seed
photons and parameters of the hot, Comptonizing plasma.

\section*{NARROW LINE SEYFERT 1 GALAXIES}


\subsection*{Energy spectra -- comparison to (broad lines) Seyfert 1 galaxies}

X--ray spectra ($E <1$ keV) of NLSy 1 contain a strong, soft component
above extrapolated hard X--ray power law (Leighly 1999b; Comastri 2000). 
This contrasts with the situation
in BLSy 1, where the presence of soft X--ray excesses was not confirmed by 
{\it BeppoSAX\/} (Matt 2000). Precise shape of the soft component is difficult
to determine, since various models seem to fit the data (almost) equally well,
if only X--ray data are used. In particular,
in some sources the best fit (for the soft component alone) is provided 
by a simple blackbody or double blackbody, while a power law is preferred
for other sources (Comastri 2000), perhaps indicating Comptonized emission.
Such Comptonized soft excess was clearly detected in multi wavelength data of
PG~1211+143 (Janiuk et al.\ 2000b), and PKS~0558--504 using
{\it XMM-Newton\/} data (O'Brien et al.\ 2000).

Spectral slopes in the {\it ASCA\/} band are softer for
NLSy 1 than for BLSy 1 (Brandt et al.\ 1997; Leighly 1999b).
Mean slope in BLSy 1 is $\approx 1.8$ while it is $\approx 2.2$ for NLSy 1
(Leighly 1999b).

The ratio of luminosities in the soft and hard X--ray components spans
a broad range of values. For example, the ratio of luminosities
$L(0.1-2\,{\rm keV})/L(2-10\,{\rm keV})$ was computed from {\it BeppoSAX\/}
data (Comastri 2000). It ranges from $\approx 2$ to $\approx 20$.

Their narrower emission lines (compared to BLSy 1) are consistent with 
the higher temperature of the soft, disk component and steeper X--ray spectrum
through photo-ionization modeling (Kuraszkiewicz et al.\ 2000 and
references therein).

\subsubsection*{Reprocessed components}

\begin{figure}
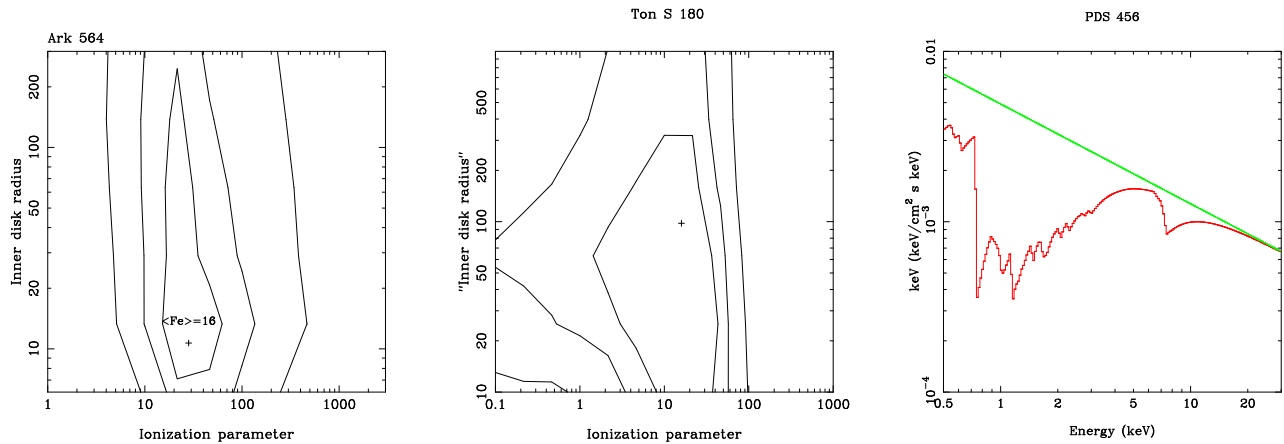

\hfil\epsfysize = 5 cm\epsfbox[18 250 600 750]{zycki_fig3a.ps}\hfil
     \epsfysize = 5 cm\epsfbox[18 250 600 750]{zycki_fig3b.ps}\hfil
     \epsfysize = 5 cm\epsfbox[18 250 600 750]{zycki_fig3c.ps}\hfil
\caption{Ionized reprocessing in NLSy1: panels a), b) show contours of 
$\chi^2$ in the $\Rin$--$\xi$ plane for two NLSy 1
galaxies, where ionized reprocessing was reported in literature.
Fitting the {\sc rel-repr} model confirms the ionization, with some
quantitative differences to the previous results: the ionization is
generally weaker in our fits. Panel c) shows the best fit model for
PDS~456, where our fits indicate that warm absorber rather than reflection
is taking place.
\label{zycki_fig:ioncont}
}
\end{figure}

A number of recent analyzes suggest that the reprocessing medium in NLSy1
is rather more strongly ionized than in BLSy1. Published results include
Ton S 180, Ark 546, PDS 456 (see reviews and references in Comastri 2000 and  
Pounds \& Vaughan 2000).
Those results are not always reliable however, since simplified models
were sometimes used. Proper model should include both the Fe \Ka\ line
and the Compton reflected continuum, with ionization effects computed
consistently for both. Also, relativistic effects, if present, affect both 
the line and the K-edge. Nevertheless, re-analysis of some of the published 
data, using the {\sc rel-repr} reprocessing model of \.{Z}ycki et al.\ (1997) 
generally
confirms those results (Janiuk et al.\ 2000a). Contour plots of $\chi^2$
(Fig.~\ref{zycki_fig:ioncont}) show however that quantitative
results on ionization and relativistic effects are  dependent upon
the model for the continuum. In one case, PDS 456, our re-analysis of 
simultaneous {\it ASCA}--{\it RXTE\/} observation suggests different 
interpretation to that previously published (Reeves et al.\ 2000). 
We find that double warm absorber gives much
better fit to the data than any model involving the reprocessed component
(Fig~\ref{zycki_fig:ioncont}).
Simple fits to {\it ASCA\/} 0.8--10 keV data 
give: \par
 1. {\sc wabs( powerlaw )}: $\hi=649/605$, $\Gamma\approx 1.4$; \par
 2. {\sc wabs*absori(powerlaw)}: $\hi=625/603$, $\Gamma\approx 1.3$, and the 
  main   improvement comes from modeling the Fe K-shell edge; \par
 3. {\sc wabs(powerlaw + rel-repr)}: $\hi= 637/603$, \par
 4. {\sc wabs*absori*absori(powerlaw)}: $\hi = 576/602$, $\Gamma\approx 2.5$ \par
 5. adding now the {\sc rel-repr} model does {\em not\/} improve the fit.\par
It is clear that only the continuum slope in fit no.~4 can be compatible with
{\it RXTE\/} data. Fitting the double warm absorber to combined 
{\it ASCA}--{\it RXTE} data (the latter in the 4--20 keV band) 
gives a good fit with $\hi=623/643$, and again adding the 
{\sc rel-repr} model does {\em not\/} improve the fit. 
The model containing a single ionized absorber and reflector, 
{\sc wabs*absori*(powerlaw + rel-repr)}, can give a fit of comparable
quality ($\hi=632/641$, i.e.\ worse by $\Delta\chi^2=9$), 
provided that the \Ka\ line is artificially suppressed (while the amplitude 
$R>1$). However, the inferred ionization of the reprocessor is 
{\em not\/} compatible with the resonant Auger destruction of the line
photons. The double warm
absorber is similar to that required to explain the MCG --6-30-15 data,
although the columns required for PDS 456 are an order of magnitude larger,
in excess of $10^{23}\,{\rm cm^{-2}}$.

\subsection*{Time variability}

NLSy 1 galaxies generally show stronger X--ray variability than BLSy 1s.
Some {\it ROSAT\/} (0.1--2 keV) observations showed dramatic variability 
with giant flares, with flux increase
by a factor of 3--5, within a $\sim$day (see Boller 2000 for recent review).
Best known examples are IRAS~13224-3809 (Boller et al.\ 1997),
PHL~1092 (Brandt et al.\ 1999) and PKS~0558-504 (Gliozzi et al.\ 2000).
However, no systematic study of {\it ROSAT\/} variability of all NLSy 1s 
has been done.

Systematic study of variability in {\it ASCA\/} data shows that indeed they
are more strongly variable than BLSy 1 in the harder X--ray band 
(Leighly 1999a), showing larger excess variance at a given luminosity.
All the three above mentioned extreme {\it ROSAT\/} sources show above-average
variability in {\it ASCA\/} (see fig.~8 in Leighly 1999a). Within
the {\it ASCA\/} band NLSy1s are more strongly variable at lower energies
than at higher energies. The ratio 
$\sigma_{\rm rms}(E<1.5\,{\rm keV})/\sigma_{\rm rms}(E>1.5\,{\rm keV})\sim 2$
(Leighly 1999a and K.~Leighly, private communication).

Systematic study of NLSy 1s variability with {\it RXTE\/} have not been
completed yet. One object for which a monitoring campaign was done, Ark 564,
does {\em not\/} show any enhanced variability compared to BLSy 1a studied
(Edelson 2000), with r.m.s.\ variability of 25\%. The r.m.s.\ variability
for another object, PG~1211+143, is somewhat higher, about 34\% 
(Janiuk et al.\ 2000b).

\subsection*{Comparison to black hole binaries}

The overall energy spectral characteristics of NLSy 1 immediately suggest that
these objects may be suppermassive analogs of Galactic BHB in high/soft state
(Pounds et al.\ 1995; Boller et al.\ 1996).
If so, they accrete at higher $\dot m$ than BLSy 1. The latter inference
is supported by the ionized reprocessed components present in their spectra
(see eq.~4 in Ross \& Fabian 1993), 
and continuum modeling (Kuraszkiewicz et al.\ 2000).

\subsubsection*{Energy spectra}

Presence of a strong, soft component is a defining feature of a soft spectral
state of BHB. The hard component may but does not need to be present. 
A soft state
is usually called Very High State, if a strong hard component is present,
and High State if the hard component is weak. Sometimes an Intermediate
State is introduced, with properties intermediate between the High State 
and Low State (Esin et al.\ 1997). 
NLSy 1 can probably correspond to any soft state.
Those with a strong hard component would correspond to Very High or 
Intermediate state, while those with a weak hard component to High State.

The {\it ASCA\/} spectral indices of NLSy 1 are similar to those
of BHB in soft states. For example, Nova Muscae 1991 had 
$\Gamma\approx 2.3$ in its IS spectrum, similarly Cyg X-1 showed spectrum
with $\Gamma\approx 2.4$ in its soft state
(note that the soft  state of Cyg X--1 seems to 
correspond of IS of soft X--ray transients, rather than HS, as the bolometric 
luminosity is virtually unchanged compared to LS).

The $F_{\rm soft}/F_{\rm hard}$ ratio in BHB can attain very large values when
a weak hard tail is present, down to $\sim 4$ in IS (from Nova Muscae 1991
data). Similarly, a range of values is seen in NLSy 1s.

A characteristic feature of hard X--ray/$\gamma$--ray emission of BHB 
in soft states is the lack of high energy cutoff in the spectra 
(Grove et al.\ 1998), indicating perhaps an inverse Compton process on
a non-thermal population of electrons. This is certainly true for
the IS, where Cyg X--1 is the best example (Gierli\'{n}ski et al.\ 1999).
Similarly, {\em GRANAT\/} spectra of Nova Muscae 1991 in VHS 
do not seem to have any cutoff below $\sim 300$ keV (Gilfanov et al.\ 1993).
GRO~J1655-40 may have been in HS or even VHS when
observed by {\it CGRO}/OSSE in 1996, with the spectrum  showing no cutoff
at least below $\sim 500$ keV (Grove et al.\ 1998). 
Unfortunately, no such data exist for NLSy 1, and studying the soft 
$\gamma$-ray spectra of those objects is not going to be feasible in near 
future.

The soft components in the soft state spectra of BHB are usually 
{\em not\/} well described by a simple blackbody, or multicolor (disk)
blackbody models (contrary to results using simplified spectral
models). Additional Comptonization of the seed, (disk)blackbody
photons is often required (i.e.\ beside the hard component extending
to $>10$ keV). This has been seen in Nova Muscae 1991 
(\.{Z}ycki et al.\ 1998a),
Cyg X--1 (Gieli\'{n}ski et al.\ 1999), GRS~1915+105 (Vilhu \& Nevalainen 1998),
GS~2000+25, XTE~J1550-564, GRO~J1655-40 (\.{Z}ycki et al., in prep.;
see also \.{Z}ycki 2000). Interestingly, similar component is seen in multi 
wavelength spectrum  of a NLSy 1 PG~1211+143 (Janiuk et al.\ 2000b),
and in the {\em XMM-Newton\/} spectrum of PKS~0558-504 (O'Brien et al.\ 2000).

The lower disk temperatures in NLSy 1 than in BHB means also that the
X--ray reprocessed component in the former objects can be expected to
be rather less strongly ionized than in corresponding states of BHB.
This indeed seems to be supported from modeling the data.

\begin{figure}
\hfil\epsfysize = 4.5 cm\epsfbox[18 520 600 700]{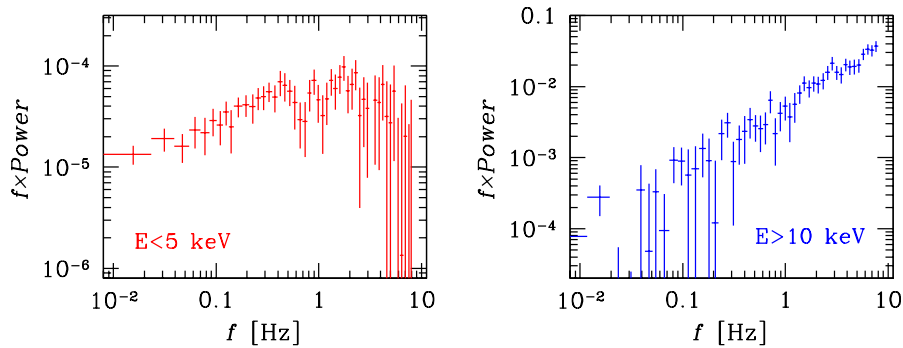}\hfil
\caption{Power spectra of Nova Muscae 1991 in VHS, on 16 Jan,
when the source spectrum was dominated by a strong, soft, thermal
component with $kT\approx 1$ keV. The soft and hard components show
very different power spectra.
\label{zycki_fig:nm16_var}}
\end{figure}

\subsubsection*{Time variability}

Again, while the energy spectra of NLSy 1s appear to be similar to their
stellar counterparts, their variability characteristics  are very different.
Generally, in BHB the soft state is the less strongly variable one
(review in van der Klis 1995), while the opposite seems to be true for 
the AGNs.

The {\it ASCA\/} data clearly show stronger variability in NLSy 1s 
than in BLSy 1s (Leighly 1999a), but it is rather unclear whether
it is the hard, Comptonized component or the soft, disk component (or both)
that  varies more strongly in NLSy 1s than in BLSy 1s. In any case,
variability of the hard, Comptonized component in NLSy 1s is not reduced
compared to BLSy 1s.

In BHB the situation is even less clear.
The hard X--ray power law is present in VHS and IS. In VHS the r.m.s.\ 
variability amplitude of the hard component is clearly lower than in 
low/hard state: $\sim 5\%$ vs. $\sim 20\%$ (Belloni et al.\ 1997 and
Takizawa et al.\ 1997 for VHS of Nova Muscae 1991).
In the IS Nova Muscae 1991 showed similarly
reduced variability of the hard component (Belloni et al.\ 1997).
However, Cyg X--1  shows basically the same r.m.s.\ variability of its hard
X--ray component in the soft state as in its hard state 
(Cui et al.\ 1997; Gilfanov et al.\ 2000a). 

The soft, disk component in NLSy 1s can show dramatic variability as 
evidenced by the {\it ROSAT\/} examples.
This component is clearly only very weakly variable in soft
states of BHB (van der Klis 1995; Belloni et al.\ 1997). 
Figure~\ref{zycki_fig:nm16_var} show PDS of Nova Muscae 1991
on 16 Jan, when the source was in VHS and its spectrum was dominated
by a strong soft component.

In the soft state of Cyg X--1 the reprocessed component was observed to 
respond to primary variability on much shorter time-scale  than in the hard 
state (Revnivtsev et al.\ 1999a; Gilfanov et al.\ 2000a). While the
situation in the hard state seems to have its correspondence in  the 
observed lack of 
reverberation signatures in some BLSy 1s, as discussed in previous
Section, no studies of variability of the reprocessed component were
performed in NLSy 1s. It is however interesting to note that the opposite
behaviour of variability as function of energy was observed in BLSy 1s
and in NLSy 1s: in NLSy 1s the harder, 7--10 keV, band varies more strongly
than the softer, 2--4 keV, band (Edelson 2000). If rapid variability
of the reprocessed component in NLSy 1s could be verified, the analogy
between BHB and NLSy 1s would certainly be more exact.

\section*{ACKNOWLEDGMENTS}

I thank B.~Czerny, C.~Done, A.~Janiuk, G.~Madejski and A.~R\'{o}\.{z}a\'{n}ska
for collaboration and discussions.
This work was supported in part by KBN grants no.\ 2P03D01816 and 2P03D01718

\end{document}